\documentstyle[preprint,aps,epsfig]{revtex}

\def\opone{\leavevmode\hbox{\small1\kern-3.8pt\normalsize1}}
 
\begin{document}

\draft

\title{Zero temperature correlations in trapped Bose-Einstein
       condensates}

\author{Walter T. Strunz}
\address{
Fachbereich Physik, Universit\"at GH Essen, 45117 Essen, Germany}
\date{\today}

\maketitle

\begin{abstract}

We introduce a family of 
correlated trial wave functions for the $N$-particle
ground state of an interacting Bose gas in a harmonic
trap. For large $N$,
the correlations lead to a relative energy decrease of
a fraction $\frac{3}{5N}$, compared to mean field
Gross-Pitaevskii theory. 
The kinetic energy 
in the weakly confining direction turns out to
be most sensitive to our correlations and, remarkably, is
{\it higher} by as much as a few per cent
for condensates with atom numbers of a
few thousand.
Thus, the predicted deviations from Gross-Pitaevskii
theory originating from ground state 
correlations might be observed in momentum
distribution measurements of small condensates.

\end{abstract}


Initiated by the first realizations of alkali Bose-Einstein 
condensates in magnetic traps a few years ago \cite{experiments}
there is now an ever growing number of experiments with atomic
condensates in laboratories all over the world. 
As a result of these efforts, a more and more detailed 
understanding of such quantum $N$-body systems is emerging. 
Properties hitherto measured are
successfully described by updated versions of many-body theories
established many years ago, see \cite{revmodphys} for a
recent review.
Very importantly, the `condensate wave function',
the most relevant object well below the transition temperature,
is determined by the mean field Gross-Pitaevskii (GP) 
equation \cite{gpequation}. 

Among future challenges remains experimental
access to the detailed nature of the $N$-particle state
of these quantum systems well below the critical temperature, exploring
new physics beyond the standard mean field description. 
This paper presents a theoretical backup of this aim,
emphasizing non-mean-field effects in {\it trapped} condensates with a 
relatively {\it small} number of atoms. Consequences of a 
correlated ground state are expected to be most prominent in 
two-body or even higher-order correlation
functions. Nevertheless, in
this paper we show that correlations may already have
a significant effect on experimentally directly
accessible quantities like the kinetic energy of
the condensate. This holds true, as we will show, whenever the particle 
number is not too large and the trap is anisotropic.

The harmonic trap potential with frequencies
$\omega_x = \omega_y$, and $\omega_z$, provides the scales 
of the problem.
We express energies in units of $\hbar\bar\omega$ where
$\bar\omega=(\omega_x\omega_y\omega_z)^{\frac{1}{3}}$, the 
corresponding oscillator ground state 
length scale is $\bar d = \sqrt{\frac{\hbar}{m\bar\omega}}$,
and thus the momentum scale $({\hbar}/{\bar d})$. In these
units the Hamilton operator,
representing the energy of $N$ interacting atoms in an
axially symmetric harmonic trap with anisotropy parameter
$\lambda = \omega_z/\omega_x$, reads
\begin{equation}\label{totalH}
H  = \sum_{i=1}^N\left\{ \frac{{\bf p}_i^2}{2} + 
\frac{1}{2} 
\left(\lambda^{-\frac{2}{3}}(x_i^2+y_i^2)+
\lambda^{\frac{4}{3}} z_i^2\right)
\right\}
+ 4\pi a \sum_{i< j} \delta({\bf r}_i - {\bf r}_j).
\end{equation}
Interactions between atoms are two-body collisions
with a strength determined by the $s-$wave scattering
length $a$ of the atoms (measured in units of the oscillator length 
$\bar d$). The problem thus depends on the three parameters
$N$, $\lambda$, and $a$.

Bose-Einstein condensation occurs when
the thermal de-Broglie wave length of the atoms becomes larger
than the mean distance between atoms. Then, loosely speaking,
a noticeable fraction of all atoms occupy the same one-particle state.
Near zero temperature, i.e. well below the critical temperature,
we therefore expect the $N$-particle state to be well approximated
by a product state $\Psi$ of one-particle states $\psi$ \cite{BlaizotRipka},
\begin{equation}\label{productstate}
\Psi({\bf r}_1,\ldots,{\bf r}_N) = \psi({\bf r}_1) \psi({\bf r}_2) 
\cdots \psi({\bf r}_N).
\end{equation}
In second-quantized language with atom field operator $\hat\psi({\bf r})$, 
the product state (\ref{productstate}) becomes
a Fock state with particle number $N$, resulting in the 
more common expressions for the manifestation of a Bose condensate, 
for instance
$\langle\hat\psi^\dagger({\bf r})\hat\psi({\bf r}')\rangle
= N\psi^*({\bf r})\psi({\bf r}')$.

The total energy of the system, $E=\langle\Psi|H|\Psi\rangle$
with Hamiltonian (\ref{totalH}), evaluated
with the product state (\ref{productstate}) is the
Gross-Pitaevskii energy functional
\begin{equation}\label{energyfunctional}
 E_{\rm GP}[\psi]  =  
 N \int\! d^3 r\; \psi^*({\bf r}) \Big[
-\frac{1}{2}\Delta +
 \frac{1}{2} 
\left(\lambda^{-\frac{2}{3}}(x^2+y^2)+
\lambda^{\frac{4}{3}} z^2\right)  
 + 2\pi a (N-1) |\psi({\bf r})|^2
\Big]\psi({\bf r}).
\end{equation}
The determination of the approximate $N$-particle ground state 
is thus reduced to finding the minimum energy wave function
of the energy functional (\ref{energyfunctional}), resulting in the
famous Gross-Pitaevskii equation \cite{gpequation} for
the condensate wave function $\psi({\bf r})$ (strictly speaking,
in GP theory the factor $(N-1)$ in front of the interaction
term is replaced by $N$; as we are interested in effects
for a fixed, finite 
number of particles $N$, we stick to $(N-1)$ throughout this paper).

By now it is experimentally established that the 
one-particle state $\psi({\bf r})$ so obtained 
does in fact well describe the
properties of the condensate. For these dilute bosonic gases
at near-zero temperature, the assumption
of a product wave function (\ref{productstate}) for the
$N$-particle state is thus a good approximation.
There is now also a rigorous proof available \cite{liebetal} 
showing that in the limit $N\rightarrow\infty$, keeping 
the product $Na$ fixed ({\it dilute}),
the true $N$-particle ground state energy is indeed given by the
the minimum of $E_{\rm GP}[\psi]$.

Nevertheless, it is clear for finite $N$ that due to two-body forces
a product wave function (\ref{productstate}) can only be an 
approximation for the $N$-particle ground state and it is worth thinking 
about possible observable deviations for not too large atom numbers $N$.

In particular, as the center-of-mass motion of the $N$-body problem
may be separated, we expect
the true ground state to be a product wave function of
center of mass ${\bf r}_{\rm cm} = \frac{1}{N}\sum_i{\bf r}_i$ and
relative coordinates, i.e.
$\Psi({\bf r}_1,\ldots,{\bf r}_N) = \psi_{\rm cm}({\bf r}_{\rm cm})\cdot
\tilde\psi({\rm relative}\;{\rm coordinates})$
with the wave function $\tilde\psi$ for the
relative motion symmetric in the ${\bf r}_i$.
This form of the wave function is clearly 
different from the product state (\ref{productstate}). 

In this paper, we are not going to separate the center-of-mass motion, 
but instead we introduce new coordinates 
\begin{equation}\label{bostrafo}
{\bf R}_i = {\bf r}_i + (C-\opone){\bf r}_{\rm cm},
\qquad\qquad (i=1,\ldots, N)
\end{equation}
that democratically incorporate the center-of-mass degree of 
freedom. 
The matrix $C$, without loss of generality, is chosen to be the
diagonal matrix
\begin{equation}\label{cormat}
C =\left(\begin{array}{ccc}
c_r & 0 & \\
0 & c_r & 0 \\
0 & 0 & c_z
\end{array}\right),
\end{equation}
where $c_r$ and $c_z$ are two positive parameters;
$C$ will play the role of an additional variational
variable and will help to lower the total energy. For a 
totally anisotropic trap one should introduce two
parameters $c_x, c_y$ instead of just $c_r$.
The Jacobian of transformation (\ref{bostrafo}) is
${\rm d}^N{\bf R} = (\det C){\rm d}^N{\bf r}$, and
notice that the choice $C=\opone$ corresponds
to the identity transformation ${\bf R}_i = {\bf r}_i$.

Apart from simplicity, transformation (\ref{bostrafo}) is
motivated by the fact that a 
wave function $\Psi({\bf r}_1,\ldots,{\bf r}_N)= \sqrt{\det C}\;\;
\Phi({\bf R}_1,\ldots,{\bf R}_N)$ is a proper bosonic wave function
whenever 
$\Phi({\bf R}_1 ,\ldots,{\bf R}_N)$ is a bosonic wave function
in ${\bf R}_i$ coordinates.
Crucially, the wave function obtained from
a {\it product} wave function in ${\bf R}_i$-coordinates
\begin{equation}\label{cpsi}
\Psi({\bf r}_1,\ldots,{\bf r}_N)=
\sqrt{\det C}\;\;\phi({\bf R}_1)\phi({\bf R}_2)\cdots
\phi({\bf R}_N)
\end{equation}
will be a {\it correlated} bosonic wave function in atom coordinates 
unless $C = \opone$, when (\ref{cpsi})
coincides with the usual product state (\ref{productstate}). 

The actual values of the transformation parameters
$c_r,c_z$ and the shape of the wave function
$\phi$ are fixed by the requirement that the total energy 
$E = \langle\Psi|H|\Psi\rangle$ evaluated 
in the space of correlated
atomic wave functions (\ref{cpsi}) should be minimal.
It is straightforward to express the energy operator 
(\ref{totalH}) of the $N$ atoms 
in new coordinates ${\bf R}_i$ and the corresponding
momenta.
Variation with respect to $C$ determines the parameters
$c_r$, $c_z$ of the
transformation (\ref{bostrafo}) with minimum energy,
for a given wave function $\phi$ in (\ref{cpsi}). We find
\begin{equation}\label{corrpars}
c_r = \lambda^{-\frac{1}{6}}\left(
\frac{\langle X^2+Y^2 \rangle}{\langle P_X^2+P_Y^2 \rangle}
\right)^{\frac{1}{4}},\;\;\;
c_z = \lambda^{\frac{1}{3}}\left(
\frac{\langle Z^2\rangle}{\langle P_Z^2\rangle}
\right)^{\frac{1}{4}},
\end{equation}
where $P_X = -i \frac{\partial}{\partial X}$, and, for instance,
$\langle Z^2\rangle = \langle \phi|Z^2|\phi\rangle$.
In deriving (\ref{corrpars}) we simplified using the fact that in 
determining the ground state $\Psi$ we may restrict
ourselves to wave functions $\phi({\bf R})$ with
$\langle{\bf R}\rangle = \langle{\bf P}\rangle = 0$.

The total energy $E=\langle\Psi|H|\Psi\rangle$
based on the correlated state (\ref{cpsi}) 
with the best matrix $C$ from (\ref{corrpars})
becomes
\begin{equation}\label{ocenergyfunctional}
E_{\rm cor}[\phi]  = 
E_{\rm GP}[\phi]  - \frac{1}{2}
\left(\lambda^{-\frac{1}{3}}\sqrt{\langle X^2+Y^2 \rangle}-
                           \sqrt{\langle P_X^2+P_Y^2\rangle}\right)^2
 - \frac{1}{2}
\left(\lambda^{\frac{2}{3}}\sqrt{\langle Z^2\rangle}-
                           \sqrt{\langle P_Z^2\rangle}\right)^2
\end{equation}
which is manifestly lower than the uncorrelated Gross-Pitaevskii energy
$E_{\rm GP}[\phi]$ (\ref{energyfunctional})
and one of the main results of this paper.

Let us first discuss result (\ref{ocenergyfunctional})
for the case of a very large number $N$ of atoms.
The difference between $E_{\rm cor}[\phi]$ and 
$E_{\rm GP}[\phi]$ is of the order of
expectation values like $\langle X^2+Y^2 \rangle$
or $\langle P_Z^2\rangle$, while the dominating term
$E_{\rm GP}[\phi]$ is $N$ times larger.
To leading order, therefore,
we may evaluate the improved total energy with
the state $\phi$ obtained from the minimum of
$E_{\rm GP}[\phi]$ alone, i.e. with the solution of the
usual GP equation.
Since, for $N\rightarrow\infty$, 
$\langle {\bf P}^2\rangle$ (`kinetic energy') 
may be neglected with respect to 
$\langle {\bf R}^2\rangle$ (`potential energy'),
we see from (\ref{ocenergyfunctional}) that in this limit,
the energy of the correlated 
state (\ref{cpsi}) may be written as 
$ E_{\rm cor}[\phi] = E_{\rm GP}[\phi] - \frac{1}{N}E_{\rm GP}^{\rm pot}
[\phi]$, where $\phi$ is the solution of the usual GP equation
and $E_{\rm GP}^{\rm pot}$ the GP potential energy. Note that
this energy decrease is possible due to the existence of the trap
potential.
Using the asymptotic Thomas-Fermi expressions
for the total GP energy
$E_{\rm GP} = \frac{5N}{14}\left(15(N-1)a\right)^{\frac{2}{5}}$
and the GP potential energy
$E_{\rm GP}^{\rm pot}[\phi] =  
              \frac{3N}{14}\left(15(N-1)a\right)^{\frac{2}{5}}$
\cite{revmodphys}, we find for the energy of our correlated 
approximate ground state
(\ref{cpsi})
\begin{equation}\label{improve}
E_{\rm cor} = (1-\frac{3}{5N})
E_{\rm GP}\;\;\;\mbox{as}\;\; N\rightarrow\infty,
\end{equation}
a small decrease for the assumed large $N$.

Far more relevant are effects of the 
correlated ground state (\ref{cpsi})
on the properties of {\it small} condensates.
Let us therefore concentrate on the full expression 
(\ref{ocenergyfunctional})
for the total energy $E_{\rm cor}[\phi]$ of the correlated state.
As in usual GP theory \cite{revmodphys,Gausstrial},
we may proceed analytically and use Gaussian trial wave functions
$\phi({\bf R})$ parameterized by two parameters $\Sigma_r$ and $\Sigma_z$ 
for the radial and axial widths (note that these are widths
in $R$ coordinates and should not be confused with the condensate
widths in physical space, the difference being of the order $N^{-1}$). 
The full energy functional (\ref{ocenergyfunctional}) becomes
a function of $\Sigma_r$ and $\Sigma_z$
only, and it is a simple numerical task to find its minimizing values.
Results based on this Gaussian approximation
are shown in the following Figures.

In Fig.1 we show the 
parameters $c_r$ (Fig. 1a) and  $c_z$ (Fig. 1b)
of our transformation
on correlated coordinates (\ref{bostrafo}) 
for the minimum 
energy correlated state (\ref{cpsi}) as a function
of particle number $N$, and for two different anisotropies $\lambda=0.04$ 
({\it cigar}), 
and $\lambda=3$ ({\it pancake}). The three corresponding different
graphs in each figure represent different interaction strengths:
$a=0.004$ (solid line), $a=0.008$ (dashed line), and $a=0.012$
(dotted line). Recall that $c_r = c_z = 1$ would correspond to
the usual uncorrelated GP ground state (\ref{productstate}). 
We clearly see that the transformation parameters 
$c_r$ and $c_z$ are larger in the weakly
confining direction, and become larger for larger interaction 
strength $a$.

Is there any hope to measure effects induced by the
correlations of the improved ground state (\ref{cpsi})?
Among the various contributions to the total energy of the
condensate, it turns out that the small {\it kinetic
energy in the weakly confining direction} is most sensitive
to our correlations and exhibits an {\it increase} of a few per
cent compared to GP theory, for small condensates.
This increase is compensated for
by a larger {\it decrease} of potential energy of the
correlated ground state, such that the total energy is indeed
lowered.

In Fig.2 we show the kinetic energy
in the weakly confining $z$-direction $E_{\rm cor}^{\rm kin,z}$
of the correlated state (\ref{cpsi}) as a function of atom number
$N$ for an anisotropy $\lambda=0.04$ ({\it cigar}).
As in Fig.1, results are shown for three
different interaction strengths,
$a=0.004$ (solid line),
$a=0.008$ (dashed line), and
$a=0.012$ (dotted line). Recall that energies are measured in
units of $\hbar\bar\omega$, i.e. the kinetic energy shown in Fig.2
is very small compared to the potential and internal
energy of the condensate.

In Fig.3 we compare the predictions of
correlated (\ref{cpsi}) and uncorrelated (\ref{productstate})
ground state for the kinetic energy $E^{\rm kin,z}$
in the weakly confining $z$-direction, again for
a cigar shaped condensate with $\lambda=0.04$
(both quantities evaluated in a Gaussian approximation).
Shown is the relative energy increase
$\Delta E^{\rm kin,z} 
= (E_{\rm cor}^{\rm kin, z}-E_{\rm GP}^{\rm kin, z})/
 E_{\rm GP}^{\rm kin, z}$, compared to GP theory.
We see that the difference may be a few
per cent, depending on interaction strength
and atom number. For larger anisotropy, the observed
difference is even larger.
Similar, yet less pronounced results are obtained for the change 
of radial kinetic energy in pancake-type condensates. 

The difference being a few per cent, high precision
momentum distribution measurements
might be able to distinguish between the predictions
of correlated and uncorrelated ground state.
In principle, the kinetic energy
could be measured indirectly via the virial relation
$E^{\rm kin, z} = E^{\rm rel} - E^{\rm pot,r}$
\cite{revmodphys,DalStri} from the knowledge of
release energy $E^{\rm rel} = E^{\rm kin}+E^{\rm int}$
\cite{release}
and radial potential energy $E^{\rm pot,r}$. In this approach,
however, the small kinetic energy is measured as the difference of
two very large quantities and it seems doubtful whether the required
precision can be achieved. 

Far more promising are direct methods to measure the momentum
distribution of the condensate, as recently established through
Bragg spectroscopy \cite{momentum}. The high precision achieved
in this experiment might be sufficient to confirm the
predicted deviations from GP theory experimentally.

Let us summarize this paper. We found a correlated 
$N$-particle wave function (\ref{cpsi}) for an interacting
Bose gas in a harmonic trap with lower
total energy than the mean field GP product state.
The existence of the trap potential
is crucial for the correlations; the energy decrease is
roughly the potential energy of {\it one} atom.
We have concentrated on effects on simple
averaged quantities, like kinetic, potential or release energy.
Consequences of our correlations will be most significant
for relatively small condensates and
a more detailed investigation of further effects
is required, focusing on
higher-order correlation functions. 

Studying the dynamics of the condensate in a correlated 
state $\Psi$ will also be of interest as excitation 
frequencies can be measured with fairly high precision.
Based on a hydrodynamic approach to superfluids,
non-mean field corrections to the frequency of elementary
excitations due to a finite gas parameter have been calculated in 
\cite{PitString} in the large $N$-limit.
Note that for dynamics, we have to replace
the correlated energy functional (\ref{ocenergyfunctional}) by the
more general expression, valid for nonvanishing 
$\langle{\bf R}\rangle$, $\langle{\bf P}\rangle$.

The correlations reported in this paper depend on the finite 
number of atoms and the existence of the trap; both conditions 
are met by current experiments. 
They are thus of very different nature than
zero-temperature non-mean field effects described by 
Bogoliubov theory ({\it quantum depletion})
\cite{revmodphys,AndBraaten}. In our approach,
Bogoliubov theory in $R$-coordinates may be introduced {\it on top}
of the correlated ground state (\ref{cpsi}).
We conclude that there is no obvious relation between the correlations 
described by our wave function (\ref{cpsi}) and the usual 
Bogoliubov quantum depletion corrections.

Last but not least,
recent progress in high precision spectroscopy of the
momentum distribution of the condensate \cite{momentum}
might lead to an experimental confirmation of our results.

I would like to thank
R. Graham, M. Fliesser, and J. Reidl for numerous
fruitful discussions. Support by the Deutsche Forschungsgemeinschaft 
through SFB 237 ``Unordnung und gro{\ss}e Fluktuationen'' is gratefully
acknowledged.

\newpage

\vspace{.2cm}
\centerline{\epsfig{figure=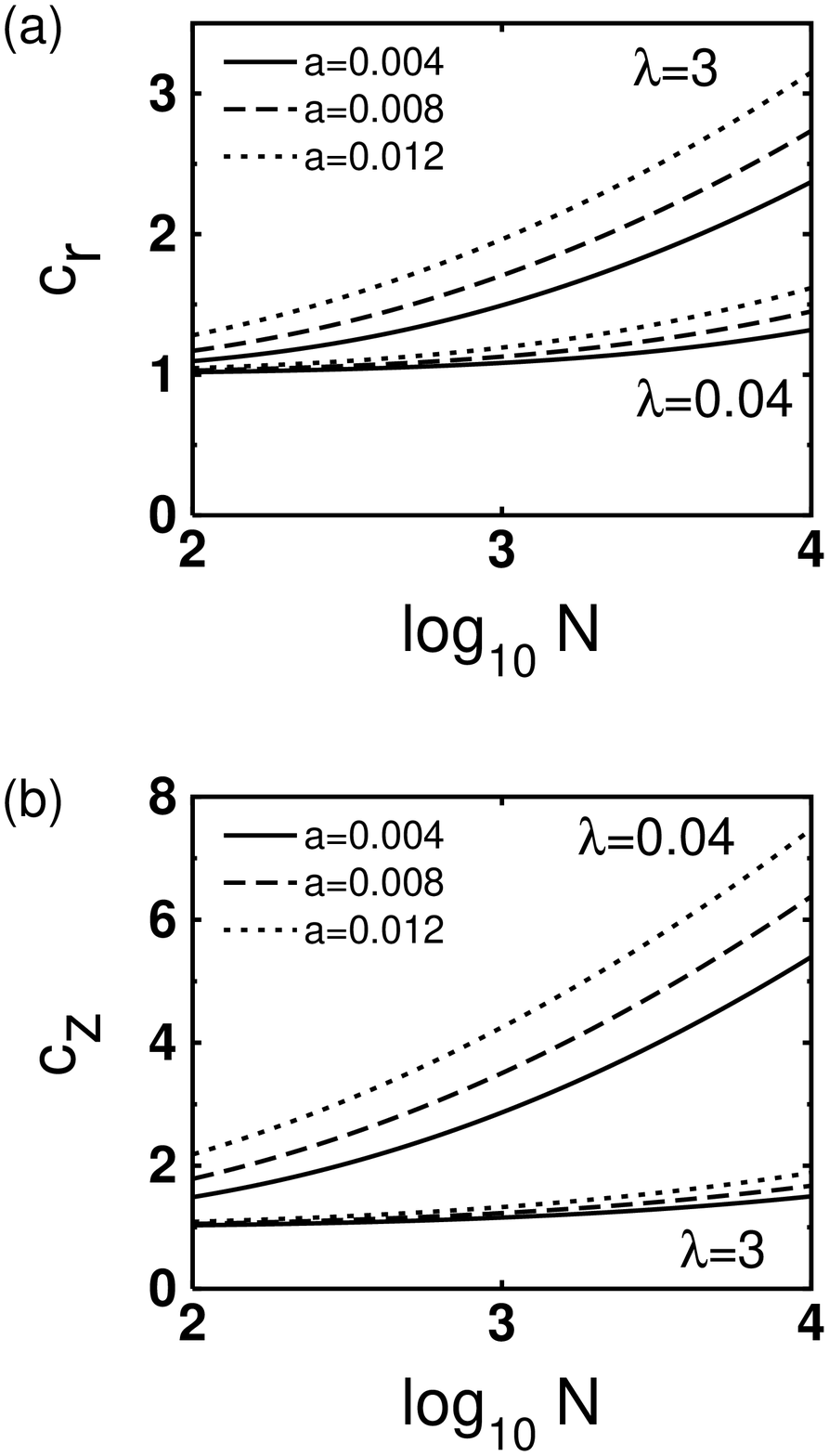,height=15cm,angle=0}}
\vspace{.2cm}

{Fig. 1. {\small{Parameters $c_r$ (Fig. 1a) and  $c_z$ (Fig. 1b)
of our transformation on correlated coordinates for the minimum energy
state $\Psi$ as a function
of particle number $N$ for two different anisotropies $\lambda=0.04$ 
({\it cigar}), 
and $\lambda=3$ ({\it pancake}). The three corresponding different
graphs in each figure represent different interaction strengths:
$a=0.004$ (solid line), $a=0.008$ (dashed line), and $a=0.012$
(dotted line). The uncorrelated case is $c_r=c_z=1$.
}}}\\

\newpage

\vspace{.2cm}
\centerline{\epsfig{figure=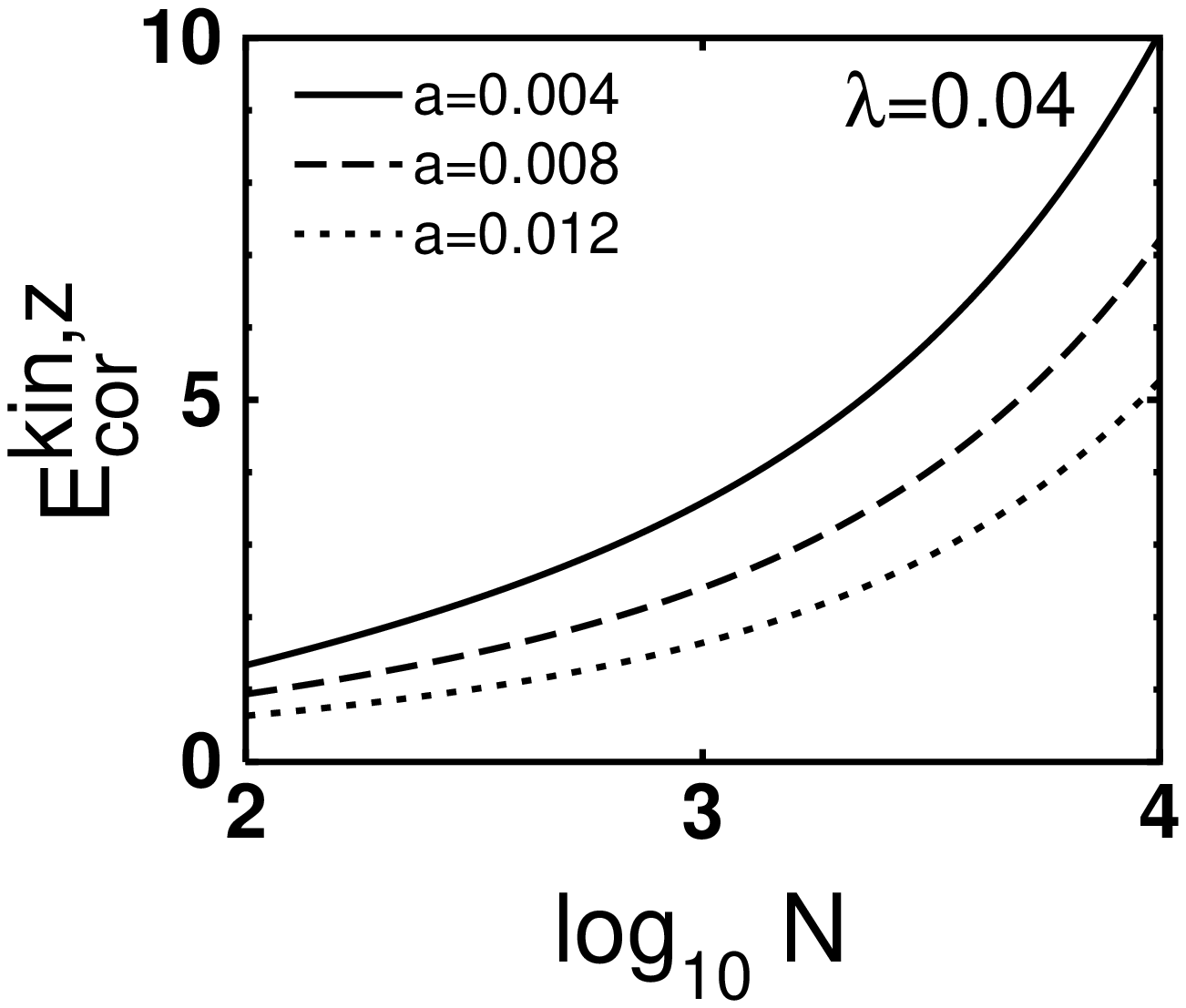,height=10cm,angle=0}}
\vspace{.2cm}

{Fig. 2. {\small{
Kinetic energy $E_{\rm cor}^{\rm pot,z}$ of the correlated state
in the weakly confining $z$-direction
as a function of atom number
$N$ for an anisotropy $\lambda=0.04$ ({\it cigar}).
As in Fig.1 we show results for three
different interaction strengths,
$a=0.004$ (solid line),
$a=0.008$ (dashed line), and
$a=0.012$ (dotted line).
}}}\\

\newpage

\vspace{.2cm}
\centerline{\epsfig{figure=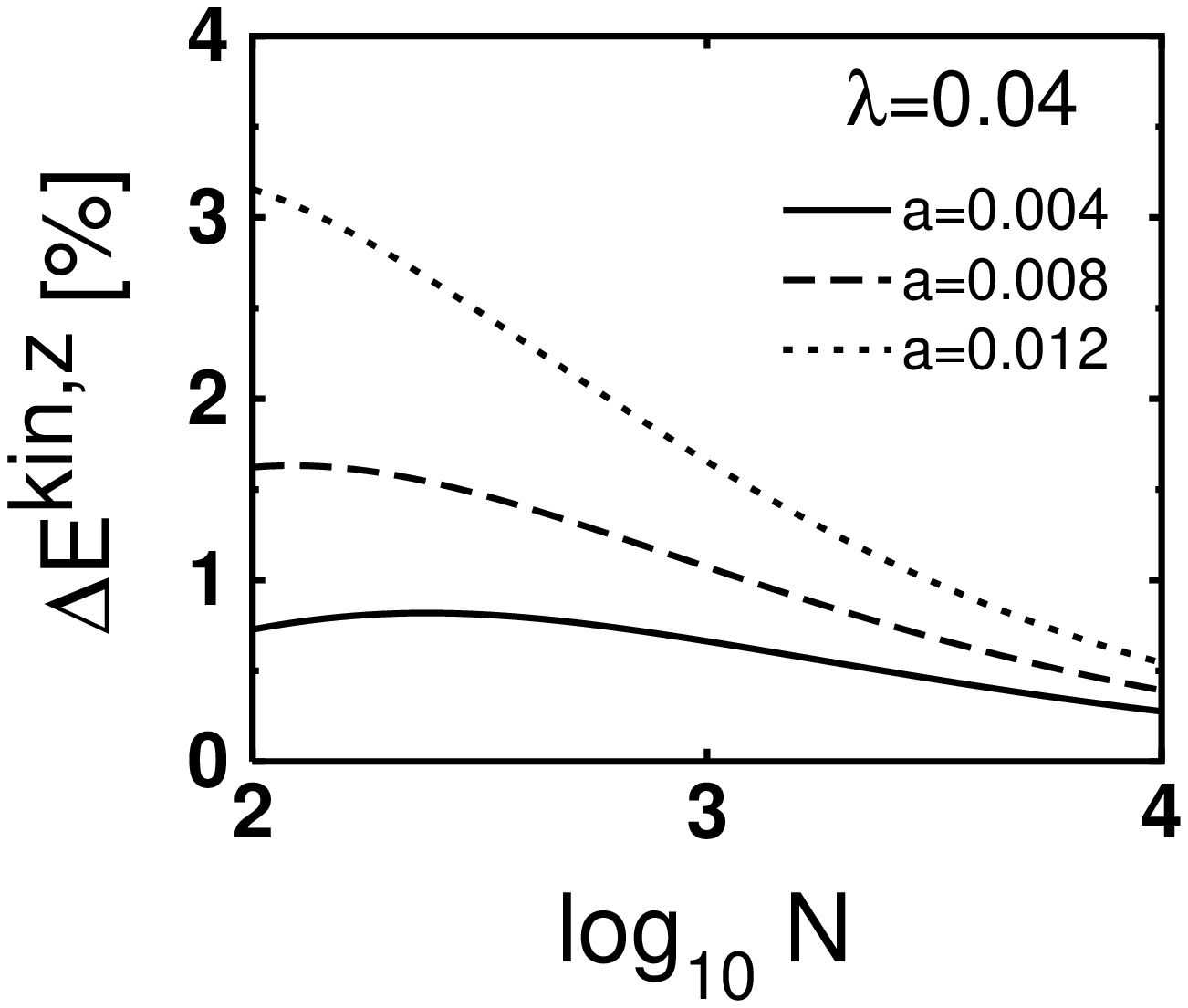,height=10cm,angle=0}}
\vspace{.2cm}

{Fig. 3. {\small{
Relative difference between the predictions for the kinetic energy 
$E^{\rm kin,z}$ 
in the weakly confining $z$-direction based on the correlated
state and the uncorrelated GP state as a function of atom 
number $N$ for an anisotropy $\lambda=0.04$ ({\it cigar}). As in 
Fig.1 and 2, we show results for three different interaction strengths,
$a=0.004$ (solid line), $a=0.008$ (dashed line), and $a=0.012$ 
(dotted line).  }}}\\

\end{document}